\documentstyle[12pt, aasms4, graphics]{article}

\begin{document}

\title{A GAMMA RAY BURST WITH A 220 MICROSECOND RISE TIME 
AND A SHARP SPECTRAL CUTOFF}

\author{Bradley E. Schaefer \altaffilmark{1} \& Katharine C. Walker
\altaffilmark{2} 
\altaffiltext{1}{schaefer@grb2.physics.yale.edu}
\altaffiltext{2}{katharine.walker@yale.edu}
}

\affil{Yale University, PO Box 208121, New Haven CT 06520-8121}

%This is single spacing
%\baselineskip 12pt
%This is double spacing
\baselineskip 24pt

\begin{abstract}

The Gamma Ray Burst GRB920229 has four extreme and unprecedented 
properties; a rise in brightness with an e-folding time scale of 
$220 \pm 30 \mu s$, a fall in brightness with an e-folding time scale of 
$400 \pm 100 \mu s$, a large change in spectral shape over a time of $768 
\mu s$, and a sharp spectral cutoff to high energies with $\Delta E/E = 18
\%$. The rapid changes occur during a spike in the light curve which was 
seen 0.164 s after the start of the burst. The spectrum has a peak 
$\nu F_{\nu}$ at 200 keV with no significant flux above 239 keV, although 
the cutoff energy shifts to less than 100 keV during the decay of 
the spike.  These numbers can be used to place severe limits on fireball 
models of bursts. The thickness of the energy production region must be
smaller than $\sim 66 km$, ejected shells must have a dispersion of the 
Lorentz factor of less than roughly $1\%$ along a particular radius, and
the angular size of the radiation emission region is of order 1
arc-minute as viewed from the burst center. The physical mechanism that
caused the sharp spectral cutoff has not been determined.
\end{abstract}


\clearpage
\section{Introduction}

The fast varying brightness has historically provided the key justification 
for the compact size of the Gamma Ray Burst (GRB) emitting region.  The idea
is that objects will not significantly change in brightness on time scales 
faster than the light crossing time of the emitter.  Significant variations
on time scales as short as $\sim 15 ms$ were apparent in early data, with
the implication that the emitter is smaller than $\sim 4500 km$, with the
further natural suggestion that neutron stars are involved somehow.

With recent 
observational discoveries (Costa et al. 1997, van Paradijs et al. 1997, 
Frail et al. 1997, Metzger et al. 1997) and theoretical models
(M\'{e}sz\'{a}ros \& 
Rees 1997), the general picture for GRBs now involves a relativistically 
expanding fireball at cosmological distances.  The relation between the 
observed time scale of brightness changes and the physical size of the 
emission region is now more complicated, and depends in detail on the 
scenario for emission (Fenimore, Madras, \& Nayakshin 1996; FMN).  

The typical upper limit on the size of the emitting region is $c\Gamma
\tau$, 
where $\Gamma$ is the Lorentz factor for the fireball expansion and $\tau$ 
is the time scale of variations(FMN).  Based on the escape of MeV and GeV
photons 
from the majority of bursts detectable by EGRET, the value of $\Gamma$ is 
generally taken to be from roughly 100 to 1000 (Harding \& Baring 1994).  
Not all bursts have high energy photons (GRB920229 in particular) and so are
not required to have the high $\Gamma$ values; but it is economical of
hypothesis 
to apply these limits to all bursts without creating special classes.

So what is the fastest time scale for GRB brightness changes?  The fast  
rise for SGR790305 (Cline et al. 1980) was a strong influence on the field 
in the 1980's, but it is now realized to apply to the separate class of 
Soft Gamma Repeaters.  Deng \& Schaefer (1997) have unsuccessfully searched 
for coherent pulsations with periods between $16 \mu s$ and $33.3 ms$, while 
Schaefer et al. (1992) has proven that microsecond  
flares do not dominate the emission as proposed by Mitrofanov (1989). With
H. A. Leder,  we have shown that the BATSE data for 20 bright bursts has no 
correlation between energies for photon pairs separated by times from 
$4 \mu s$ 
to 10 ms.  Bhat et al. (1992) reported on a burst with a total duration
of 8 ms, although the claimed 0.2 ms duration flare visible in only 
one BATSE module has problems since it is only a $3-\sigma$ detection and 
it is not visible in three other detectors which should have shown the flare.  
Nevertheless, this event and others in the BATSE catalog (Fishman et al. 1994) 
with durations as short as 34 ms, show that the time scales for 
significant variations must be as short as a few milliseconds.

We have been systematically searching through the BATSE data to find the 
 fastest time scale of variation.  The fastest time scale we found is for 
GRB920229, which has a rise time of $220 \mu s$ and a fall time of $400
\mu s$(Section 2).  During the fast fall, the burst spectrum changed
significantly 
on a time scale of $768 \mu s$ (Section 3).  In addition, this same burst
has a unique spectrum, with a sharp spectral cutoff (Section 4).  We show that 
these properties can strongly limit the fireball properties (Section 5).

\section{Fast Brightness Variations} 

Our search for rapid variability was made with the algorithm of Giles (1997), 
which is designed for the detection of infrequent rapid flares in a light 
curve.  We used data from the BATSE Large Area Detectors (LADs), each 
module of which has a frontal area of $2025 cm^{2}$.  Our variability search 
has focused on 20 of the brightest BATSE bursts, since only these will have 
the number of photons required to detect submillisecond variations.

We have been using the TTE data type from the BATSE LADs, which has each 
photon individually tagged as to time and energy.  The time is given to a 
resolution of $2 \mu s$, which is good for rapid variability studies.  The 
energy is given to within one of four discriminator channels with energy 
ranges of roughly 25-50, 50-100, 100-300, and $>300$ keV (channels 1
through 4 respectively).  The pulse pile-up time is $0.25 \mu s$ and the dead 
time is $0.13 \mu s$.  The on-board memory will store up to 32768 individual 
photons, which usually allows for only the first few seconds of the time 
history to be so recorded, often including substantial portions of the 
pretrigger light curve.

The fastest variability was discovered for GRB920229 (BATSE trigger
\#1453). The overall light curve (Figure 1) is unusual, consisting of 
a smooth symmetric 
peak followed by a much narrower spike.  The burst has a short total duration
 of 0.19 seconds.  The spike has a total duration of roughly 0.0044 s and 
occurs 0.164 seconds after the start of the burst.  Figure 2 shows the time 
history of this spike in the four energy channels.

The rise time of the spike is fast.  The count rate goes from near zero  
counts to a maximum in $512 \mu s$.  The e-folding time scale
($\tau_{rise}$) depends 
on the channel, the time range for the fit, and the size of the time bin.  
The fitted $\tau_{rise}$ values (over the fastest 
changing $512 \mu s$ time interval in the $64 \mu s$ binned light curve) are 
$190 \pm 40 \mu s$ for channel 1, $250 \pm 50 \mu s$ for channel 2, $240
\pm 50 \mu s$ 
for channel 3, and $230 \pm 30 \mu s$ for the sum of channels 1, 2, and
3. By slight changes in the time range for the fit, a time scale 
as short as $176 \pm 27 \mu s$ can be found, while the inclusion of time 
intervals substantially before the rise or after the peak will yield longer 
time scales.  For the $128 \mu s$ binned light curve for the fastest 
changing $512 \mu s$ time interval, the fitted $\tau_{rise}$ values are 
$190 \pm 50 \mu s$ for channel 1, $240 \pm 50 \mu s$ for channel 2, 
$220 \pm 50 \mu s$ for channel 3, and $220 \pm 30 \mu s$ for the sum of 
channels 1, 2, and 3.  We 
will adopt a rise time equal to $220 \pm 30 \mu s$ as conservatively 
representing the range of fitted values.  
 
The spike also has a fast e-folding fall time ($\tau_{fall}$).  The
fastest time 
scale is in channel 3.  The fitted e-folding time scale of the fall is 
$440 \pm 130 \mu s$ over 11 time bins in the $64 \mu s$ light curve
starting 
at the highest point in the spike.  In the $128 \mu s$ light curve,
$\tau_{fall}$ is 
$390 \pm 70  \mu s$ over the $1024 \mu s$ time interval after the highest 
point in the spike.  If the later half of this fall is fitted separately 
(hence missing the small interval of brightening in the middle), we find 
$\tau_{fall} = 150 \pm 70 \mu s$ and $140 \pm 60 \mu s$ for the $64 \mu s$ 
and $128 \mu s$ time bins.  The sharp drop in channel 2 starts
approximately $256 \mu s$ after the sharp drop in channel 3.  Channel 2
has a slower fall rate 
than in channel 3, its overall e-folding time scale is $760 \pm 170 \mu s$ 
while its fastest falling $512 \mu s$ time interval has $\tau_{fall} = 
310 \pm 120 \mu s$. 
The sharp drop in channel 1 starts approximately $512 \mu s$ after the
sharp drop in channel 3.  The value of $\tau_{fall}$ varies around $500
\mu s$ depending on 
the precise choice of fitted time interval, although the steepest time 
interval has $\tau_{fall} = 420 \pm 160 \mu s$.  From this analysis, we
can be confident that the decay time scale is at least as short as 
$400 \pm 100 \mu s$, 
while some time intervals may have $\tau_{fall}$ significantly shorter.

\section{Fast Spectral Variations}

The delay between the start time for the spike's sharp drop as well as the 
differences in decay time scales between the channels implies that the 
spectral shape is rapidly softening.  For most of the time during the spike, 
channel 3 has the most counts while channel 4 has essentially zero counts, 
thus indicating some spectral cutoff or break at energies just below 300 keV. 
For the tail of the rapid decay, channel 2 has the most counts while 
channels 3 and 4 have few counts, thus indicating some spectral cutoff 
or break at energies just below 100 keV.  The break energy has shifted 
significantly on a fast time scale.
 
A plausible explanation for this shift relates to the scattering of burst 
photons off the Earth's atmosphere.  Bhat et al. (1992) showed that such 
reflections are softer and delayed by $\sim 2500 \mu s$.  For the fall of 
the sharp spike in GRB920229, a soft and delayed echo might be able to 
produce the fast spectral change.  However, we do not think that 
this effect is the cause.  First, LAD detectors 1 and 5 have large angles 
away from the center of the Earth ($101^{o}$ and $122^{o}$), so only
distant parts 
of the Earth's limb will be obliquely visible to the detectors.  Second, 
LAD detector 0 (which points $79^{o}$ away from the burst yet only
$30^{o}$ away 
from the center of the Earth) sees 40 counts above background in all four 
channels over the entire burst while 41 counts are expected by 
simple scaling from LAD detectors 1 and 5, thus showing that the reflected  
flux is small. Third, LAD detector 2 (which is hidden from the burst yet 
is pointing $57^{o}$ from the Earth's center) sees only 12 counts above 
background in all four channels during the entire time of the burst, showing 
that the Earth scattering is small.

To be quantitative about the spectral changes, we have defined a hardness 
ratio as the counts in channel 3 divided by the counts in channels 1 and 2. 
The flux during the spike has a hardness ratio of typically 0.5, as compared 
to hardness ratios of $\sim 0.2$ before and after the spike when the flux is 
weak.  The hardness ratio varies from $0.60 \pm 0.12$ to $0.12 \pm 0.05$
between time bins separated by $768 \mu s$ during the decay of the spike. 
This is the fastest known time scale for spectral changes
($\tau_{spectrum}$). 
 
\section{Spectrum}

The channel 4 light curve (for both the spike and the entire burst) shows 
near zero flux.  This is surprising since the burst is brightest in channel 
3. (The background subtracted photon count over the entire burst for 
channels 1, 2, 3, and 4 are 730, 1630, 2490, and 120 photons.)  By itself, 
this shows that the spectrum for GRB920229 must have a sharp spectral 
cutoff just below the energy boundary between channels 3 and 4.  This 
conclusion does not depend on details of the background subtraction and 
spectral deconvolution.  The only observational questions are the energy 
and sharpness of the spectral break.

To measure the properties of this spectral break, we have extracted the  
spectrum from the BATSE data.  In particular, we have used the count 
spectra from LAD detectors 1 and 5, which are facing at angles of $23^{o}$ 
and $47^{o}$ from the burst direction. (The spectroscopy detectors have
too small 
an area to catch a usable number of photons, while the other triggered LAD 
modules have a large angle towards the burst direction.)  We use the HERB 
data type, which has 128 channel energy resolution and $5.1 \mu s$ dead
time 
at 100 keV.  We used the spectra which record the first 1.92 seconds after 
the BATSE trigger, so that the HERB spectra from the two detectors would 
cover identical time ranges.  The BATSE trigger time is 0.053 s after the 
start of the burst (as recorded in the pretrigger TTE data), so roughly
30\% of the burst flux is not included. The hardness ratios do not change 
significantly throughout the burst (until the end of the spike, see Section 3),
 so the exclusion of some fraction of the burst flux is unlikely to affect 
the measured spectral shape.

Our analysis procedures are given in detail in the BATSE spectroscopy catalog
 (Schaefer et al. 1994).  We have calculated the background at the time of
the burst on a channel-by-channel basis, by averaging the HERB spectra taken 
from 3.2 to 80.7 seconds after the trigger. Corrections due to Earth 
scattering of the burst photons are small since the two LAD detectors are 
pointing $101 ^{o}$ and $122 ^{o}$ from the center of our planet. The
spectra from 
both detectors are fully consistent with each other.  These individual 
spectra have been combined together and then individual energy channels have 
been binned together by the procedures in Bromm \& Schaefer (1998).  The
usual $dN/dE$ photon spectrum (with units $photon \cdot s^{-1}
\cdot cm^{-2}
keV^{-1}$) has
been multiplied by $(E/100keV)^{2}$, to get a quantity proportional to
$\nu F_{\nu}$. This spectrum runs from 22 keV to over 1 MeV.  The final
spectrum is presented in Figure 3 and Table 1.

We have performed a variety of model fits to the count spectra.  A sharply 
broken power law provides a good fit, with the $\nu F_{\nu}$ power law
indices of $+0.7 \pm 0.1$ for energies below 200 keV and $-3.5 \pm 1.0$ 
for energies above 200 keV.  The fit to the 'GRB function' (Band et al.
1993) converges 
with the break at high energy, so that the fit is really that of a power law 
times an exponential.  The best fit $\nu F_{\nu}$ power law index is $1.2
\pm 0.1$ and the characteristic energy for the exponential is $110 \pm 5
keV$. The shocked synchrotron model (Tavani 1996) does not have a sharp 
cutoff, even with the $\delta$ 
value being allowed to become large, with the best fits approaching a 
power law (with $\nu F_{\nu}$ index held at 1.33) times an exponential
(with characteristic energy of $130 \pm 10 keV$).  Both the GRB function 
and the synchrotron model systematically underestimate the flux for all bins 
from 145-222 keV and systematically overestimate the flux for all bins
$>222 keV$. Both predict that the burst photons in channel 4 should be $\sim 
1500$, whereas only 120 are observed. Thus, neither the GRB function nor 
the synchrotron model are consistent with the sharp observed break.

The peak $\nu F_{\nu}$ is close to 200 keV.  Above this peak, the flux
drops rapidly, 
with the 223-239 keV bin having half the peak $\nu F_{\nu}$ value and only
a $1.9-\sigma$ excess above background. The next higher bin is a factor of
8 below 
the peak and is only a $0.4-\sigma$ detection of flux.  All higher bins
have no 
significant flux.  So GRB920229 goes from its peak $\nu F_{\nu}$ to near
zero flux from roughly 200 to 239 keV.  This corresponds to a $\Delta E/E$ 
value of $18\%$. This is by far the sharpest known spectral break or
continuum feature (cf. Schaefer et al. 1993). The only sharper claimed 
features in GRB spectra are the proposed cyclotron lines (Murakami et al. 
1988).

\section{Implications}

GRB920229 has four extreme properties, far beyond any previous known 
example.  It has the fastest rise time ($\tau_{rise} =220 \pm 30 \mu s$),
the fastest 
fall time ($\tau_{fall} = 400 \pm 100 \mu s$), the fastest spectral
variability time scale ($\tau_{spectrum} = 768  \mu s$), and the sharpest 
spectral continuum feature ($\Delta E/E$ = 18 \%). These values provide a 
challenge to the standard fireball models of bursts.

The rise in brightness can be smeared in time for any of several kinematic 
effects.  We can analyze each effect by itself, assuming that the other 
effects are negligible, so as to produce conservative limits.  The details 
will depend on the scenario under consideration. Here, we will consider 
three general cases; the external shock model, the internal shock model, 
and a collimated jet model.

In the external shock model, some central energy release sends out a single 
shell of ejecta at relativistic velocities (with Lorentz factor, $\Gamma$,
from 100 to 1,000) which then impacts onto an ambient cloud converting the 
shell's kinetic energy into radiant energy.  A near zero rise time can be 
observed if the cloud is physically small and if the shell is physically thin.

If the cloud has a physical thickness along the line of sight, then a 
measurable rise time will result. This thickness can result either because 
the cloud has lower density towards the oncoming shell (i.e., the cloud has 
a fuzzy edge) or because the shape of the cloud results in some regions being 
hit first with the remainder being hit later (i.e., the cloud is not planar).  
The rise in brightness will start when the shell first hits the leading edge 
of the cloud with a peak corresponding to when the shell passes over the 
center of mass of the cloud. The observed rise time in brightness will
correspond to either the scale height for the cloud density or to a 
characteristic distance scale for the 
cloud structure.  The shell is moving very close to light speed, so the 
observed rise time provides a stringent limit on the size scale of the cloud 
as $2\Gamma ^{2} c \tau_{rise}$(FMN).  For GRB920229 (with $\tau_{rise} =
220 \mu s$) and $\Gamma < 1000$, the limit is 
that the characteristic distance scale of the cloud along the line of sight 
must be less than 0.9 Astronomical Units. For $\Gamma = 100$, the cloud
must be roughly the size of our Sun.

If the cloud has a physical thickness perpendicular to the line of sight, 
then a measurable rise time will result.  The reason is that the farther off 
the line of sight a photon is produced, the longer the path length for the 
photon to reach Earth.  The minimum rise time occurs when the cloud is 
centered along the line of sight, so this case will be assumed to provide the 
most conservative limits.  Since the shell is expanding at nearly the speed of 
light, the observed delay depends only on the radius of the shell at the 
time of impact with the cloud (R) and the angular radius of the cloud subtended
 from the burst site ($\Theta_{cloud}$).  The rise time will be close to
$R \Theta_{cloud}^{2}/2c$(FMN). 
The shell has been expanding for at least the time from the start of the burst 
until the time of the spike ($T_{spike}$), so $R > 2c
\Gamma^{2} \cdot T_{spike}$(FMN).  Then, 
$\Theta_{cloud} < (\tau_{rise}/T_{spike})^{0.5}/\Gamma$.  For the
GRB920229
spike, $\tau_{rise} = 220 \mu s$, $T_{spike} = 0.164 s$, and $\Gamma > 
100$, we get that $\Theta_{cloud}$ must be less than 
$4 \times 10^{-4}$ radians or 1.3 arc-minutes.  Due to self shadowing,
Earth can only see a 'cap' of the shell which subtends an angle
$\Theta_{cap} = \Gamma^{-1}$, so the cloud can 
only occupy roughly 0.13\% of the cap region at most. Fenimore et al.
(1998) have examined the smoothness of burst light curves to reach a
similar conclusion that the surface filling factor is typically $\sim 0.5
\%$. 

If the shell has a physical thickness, then a measurable rise time will  
result. Such a thickness could arise either because the energy release volume 
is large or because the ejecta has a dispersion of velocities.  For the rise 
time to be $220 \mu s$, the characteristic thickness of the energy
release region must be less 
than roughly 66 km(FMN).  Even a slight dispersion of velocities will
result in a 
precursor of fast ejecta significantly preceding the main mass of the shell 
hitting the cloud, with a long rise time.  Let $<\Gamma>$ be the Lorentz 
factor for the densest layer of the shell, with $\Delta \Gamma$ the
difference in Lorentz factor between $<\Gamma>$ and the leading edge of the 
shell. To allow
for the observed rise time, the fractional dispersion in Lorentz factors 
within the shell must be less than $\tau_{rise}/2T_{spike}$ along the
radius vector. For the spike 
in GRB920229, the shell must have $\Delta \Gamma/<\Gamma>$ less than
0.07\%.  In the absence of a significantly dense 
medium to contain the expansion, we know of no means to enforce such a 
constant velocity.

In the internal shock model, some central engine sends out multiple shells 
at relativistic velocities, and these shells collide with each other releasing 
their kinetic energy as radiation.  A near zero rise time would be visible if 
the shells were physically thin and do not cover a large solid angle.

If the energy release region has a significant characteristic size, then a 
measurable rise time would result.  That is, a large emission region would  
produce a shell with a leading edge that has a thickness comparable to the 
region's size, and then produce a rise in brightness approximately equal to 
the light travel time across the source region.  For GRB920229, the source 
region must then be smaller than $\sim 66 km$.

If one of the colliding shells has a dispersion of velocities, then a 
measurable rise time would result.  That is, the fast ejecta would strike the 
outside shell first while the densest part of the shell would hit only later. 
 The rise time depends on the time delay between the emission of the two 
shells ($\Delta T$), the ratio of Lorentz factors between the outer shell
and the fast 
material in the inner shell ($L_{of}$), and the ratio of Lorentz factors
between the outer shell and the densest material in the inner shell ($L_{od}$).
We must have $L_{of} < L_{od} < 1$.  The rise time will be
$\Delta T(L_{od}^{2} - L_{of}^{2})/[(1-L_{od}^{2})(1-L_{of}^{2})]$. If 
$L_{of}$ is not close to $L_{od}$, then the rise time will be of order
$\Delta T$. If the value 
of $\Delta T$ is of order $220 \mu s$, then the shells will collide within
a few hundred 
kilometers of the central engine unless the shell velocities are closely 
matched.  Alternatively, $\Delta T$ can be substantially larger (say,
comparable to 
the burst duration) if $L_{od} \approx L_{of}$.  For GRB920229 with
$\tau_{rise} = 220 \mu s$, if 
$\Delta T \sim 0.17$ seconds and $L_{of} = 0.2$, then the dispersion of
Lorentz factors in 
the shell must be less that 1\%.  For any alternative, the ejecta
velocities will have to be finely tuned.

If the colliding shells subtend a significant solid angle (as viewed from the 
site of the central engine), then a measurable rise time would result. 
Portions of the collision that are off the line of sight will have a delay due 
to the longer path length that the photons must travel to reach Earth.  
The limits are the same as derived for the external shock model, except in 
this case we cannot constrain the radius of the shells at the time 
of the collision. Thus, no formal constraints on the solid angle of the 
collision region arise from this effect.  That is, if two shells are ejected 
such that they collide at a great distance from the central engine, then the 
delay for the off axis photons can be made small.  Any such attempt to make R 
very large then forces $\Delta T$ to increase so that the rise time will
nonetheless 
be long unless the shell velocities are finely tuned. However, for R values 
comparable to those deduced from the time of first visible gamma radiation, 
limits on the subtended angle for the collision region will be of order an 
arc-minute.  

Jet models would have the relativistic outflow confined to a small solid 
angle, with the radiation produced by either external or internal shock. 
In either case, the limits derived above are still applicable.  (Models 
where the jets sweep across either the Earth or some cloud have already been 
rejected due to the strong asymmetry of rise and decay times; Nemiroff et al. 
1994.)  Of particular interest is the very narrow collimation of the beam 
($< 1$ arc-minute) in the case where the emission is produced by a
collision 
with a cloud.  Even if the radiation is produced by shocks internal to the 
jet, comparable collimation constraints are required for reasonable R values.

What physical mechanism produces the sharp spectral cutoff? Within the 
framework of the fireball model for GRBs, we are seeing flux from particles 
expanding towards Earth at relativistic velocities. So in the rest frame of 
the emitting particles, the spectral cutoff is actually at a substantially 
lower energy. For typical estimates that $\Gamma$ is 100-1000, the cutoff 
must be from 0.2 to 2.0 keV. Electron-positron opacity effects are not
sharp enough and are always at higher energy; cf. Baring 1990. Photoelectric 
absorption would require unionized matter and would cutoff to 
lower energies. The shocked synchrotron model cuts off as an exponential in 
the absence of a hard tail, and so is already rejected by the model fits.
The mechanism must allow for a shift in the observed cutoff energy, 
as seen during the decay of the spike. If a single physical process can be 
identified, then the observed cutoff characteristics could yield unique 
information on properties such as temperature, optical depth, or $\Gamma$.
 
The sharp spectral cutoff can constrain the velocity dispersion within the 
expanding shells.  A spread of velocities would cause the cutoff for each 
volume element to be boosted by a different Lorentz factor.  For the 
observed $\Delta E/<E> = 18\%$ , the $\Delta \Gamma/<\Gamma>$ value must
be less than 18\%.

The opening paragraph of this section summarized the extreme properties  
of GRB920229, and these place severe constraints on burst models.  The
thickness of the energy production region must be smaller than $\sim 66
km$ along the line of sight, ejected
shells must have a dispersion of the Lorentz factor of less than roughly 1\%, 
and the angular size of the radiation emission region is of order 1 arc-minute 
as viewed from the burst center. The physical mechanism for the spectral 
cutoff (at an energy of 200 keV with no blue shift or 0.2-2.0 keV in the 
rest frame of the shell) has not been determined.

We thank E. E. Fenimore for key discussions.

\begin{table}
\begin{center}
\begin{tabular}{|c|c|}
\hline
Bin energy &  $\nu F_{\nu} \times 1000$ \\
\hline
22 - 33   & $0.63  \pm 0.19$ \\
33 - 45   & $0.65  \pm 0.26$ \\
45 - 58   & $1.60  \pm 0.29$ \\
58 - 72   & $1.77  \pm 0.32$ \\
72 - 86   & $1.75  \pm 0.37$ \\
86 - 100  & $2.37  \pm 0.42$ \\
100 - 115 & $2.80  \pm 0.47$ \\
115 - 130 & $2.70  \pm 0.50$ \\
130 - 145 & $2.47  \pm 0.55$ \\
145 - 160 & $3.68  \pm 0.67$ \\
160 - 176 & $3.15  \pm 0.76$ \\
176 - 191 & $3.73  \pm 0.89$ \\
191 - 207 & $4.26  \pm 1.03$ \\
207 - 223 & $3.32  \pm 1.15$ \\
223 - 239 & $2.34  \pm 1.25$ \\
239 - 255 & $0.52  \pm 1.38$ \\
255 - 271 & $0.61  \pm 1.59$ \\
271 - 287 & $0.74  \pm 1.82$ \\
287 - 325 & $1.49  \pm 1.49$ \\
325 - 457 & $0.10  \pm 1.25$ \\
457 - 592 & $-0.97 \pm 2.97$ \\
\hline
\end{tabular}
\caption{GRB920229 spectrum.}
\end{center}
\end{table}

\begin{figure}
\begin{center}
\resizebox{12cm}{10cm}{\includegraphics{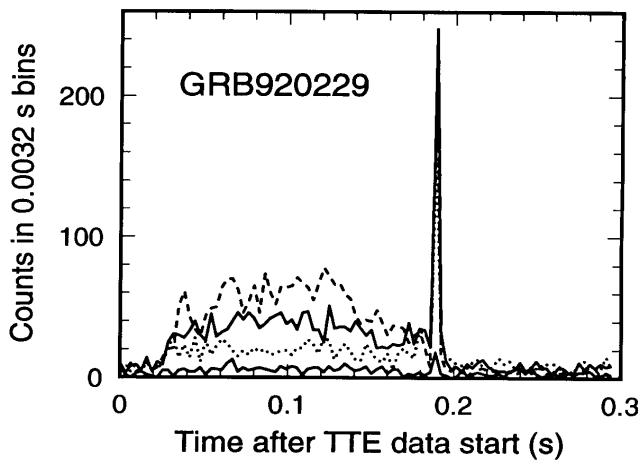}}
\caption{Light curve for GRB920229. BATSE trigger \#1453 is a short burst 
which ends in a bright spike.  This plot shows each of the four channels 
from TTE data with a time resolution of $3200 \mu s$. The time axis is 
in seconds with respect to the BATSE trigger time.  An expanded view of 
the spike's light curve is presented in Figure 2.  The relative counts in 
the channels do not change throughout the burst.  The curves, in ascending 
order at 0.1 s, are channels 4, 1, 2, and 3.  Note that channel 3 (for 
photons from 100 to 300 keV) has the most counts while channel 4 (for 
photons with energies $> 300 keV$) has virtually zero flux throughout the 
burst, thus indicating the existence of a sharp spectral break somewhere 
just below 300 keV.}
\end{center}
\end{figure}

\begin{figure}
\begin{center}
\resizebox{12cm}{10cm}{\includegraphics{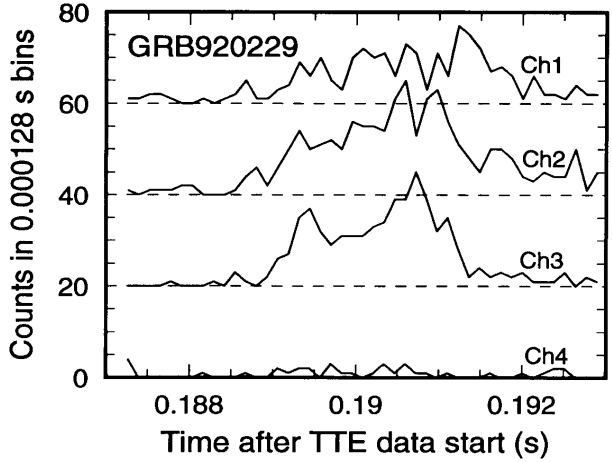}}
\caption{The spike in GRB920229. The light curve for the spike is presented 
with $128 \mu s$ time resolution for each of the four energy channels with an 
offset of 20 counts between each channel for clarity. Channel 4 is seen to 
have no significant flux above background.  The rapid brightening centered 
on a time of 0.189 s has trise $220 \pm 130 \mu s$.  The fading centered on 
time 0.191s has $\tau_{fall} = 400 \pm 100 \mu s$.  The fall commences
later and 
decays slower for the low energy channels, resulting in a significant 
spectral change between times of 0.1907 and 0.1915 s.}
\end{center}
\end{figure}

\begin{figure}
\begin{center}
\resizebox{12cm}{10cm}{\includegraphics{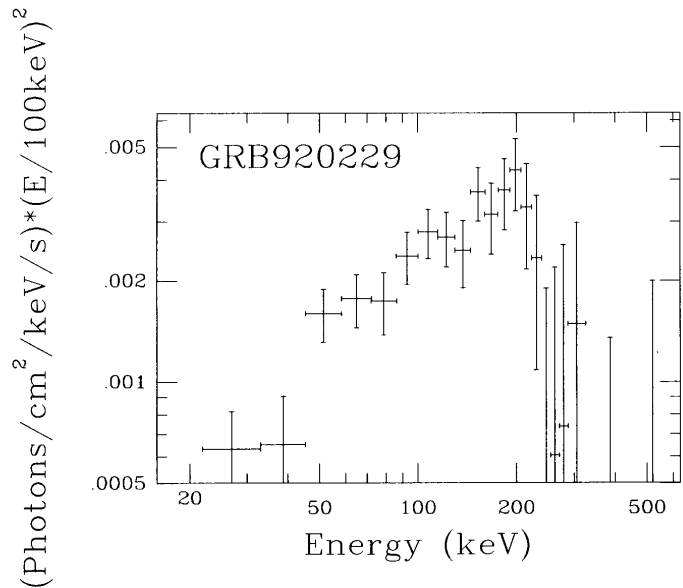}}
\caption{Spectrum of GRB920229. This photon spectrum is the combined result 
of LAD detectors 1 and 5, and covers the entire burst after the BATSE 
trigger time.  The peak is close to 200 keV, with a normal spectral slope 
to lower energies.  Above 200 keV, the flux drops steeply, with no 
significant flux above 239 keV.  This spectral cutoff (with $\Delta E/<E> = 18
\%$) is by far the sharpest known continuum feature for bursts, and hence might 
carry valuable physical information if the cause of the cutoff were known.}
\end{center}
\end{figure}

\end{document}